\documentclass[12pt,letterpaper]{article} \usepackage[latin1] {inputenc} 
\usepackage{amsmath} \usepackage{amsfonts} \usepackage{amssymb} 
\usepackage{amscd}
\begin{document}
\title{A Variational Formulation of  Symplectic Noncommutative Mechanics}
\author{Ignacio Cortese\footnote{nachoc@nucleares.unam.mx}\,\, and  J. Antonio Garc\'\i a\footnote{ garcia@nucleares.unam.mx}\\
\em Instituto de Ciencias Nucleares, \\
\em Univesidad Nacional Aut\'onoma de M\'exico\\
\em Apartado Postal 70-543, M\'exico D.F., M\'exico}
\maketitle
\abstract{The standard lore in noncommutative physics is the use of first order variational description of a dynamical system to probe the space noncommutativity and its consequences in the dynamics in phase space. As the ultimate goal is to understand the inherent space noncommutativity we propose a variational principle for  noncommutative dynamical systems in configuration space, based on results of our previous work \cite{CG1}. We hope that this variational formulation in  configuration space can be of help to elucidate the definition of some global and dynamical properties of classical and quantum noncommutative space. 
}

\section{Introduction}

From the point of view of classical dynamics the recent interest in noncommutative dynamical systems \cite{Horvathy1}  boils down to the study of generalized Hamiltonian dynamics where the Poisson structure involved could be a complicated function of phase space coordinates and momenta. Of course this class of dynamical structures are well known starting from the Hamiltonian formulation of Euler equations for the rigid body dynamics to models of hydrodynamics and integrable non linear systems of partial differential equations \cite{String}. It is quite surprising that this venerable subject is still producing new results and insights in recent research areas as condensed matter physics \cite{Horvathy2}, quantum Hall effect \cite{QH},  Snyder space \cite{Snyder}, and double special relativity  \cite{DSR} models. When quantized, these models could have new physics at some very high energy
scale.  It is still not clear if this perspective will be of some utility to revel some aspects of quantum gravity, nevertheless it could help for a better understanding of the more complicated noncommutative field theories whose aim is to explain new physics near or beyond Planck scale. A point that is not sufficiently  stressed in recent literature about noncommutative physics is that the usual way of doing physics through dynamical models consist in construction of the interaction based on physical grounds and symmetry principles. In many circumstances the model can be brought to a form that is suited for a variational formulation in a Lagrangian or Hamiltonian version. In Hamiltonian form the interaction can be handled in the definition of the Hamiltonian function itself or can be taken into account using a simple Hamiltonian and a deformed Poisson bracket that implement the interaction in the equations of motion. Of course the description is completely covariant in phase space due to the Darboux theorem. But in noncommutative physics the argument is reversed. Now the relevant interaction information lie in ONE system of coordinates where the Poisson bracket have a peculiar distinctive form, implementing space noncommutativity. Then we proceed to explore the consequences of this peculiar Poisson brackets on the dynamics. Needless to say that the usual way of working the system properties is to implement standard Darboux coordinates and read off the corresponding interaction Hamiltonian. A  drawback of this procedure is that in many cases the interaction Hamiltonian depends in a complicated way of the momenta. As a consequence, the elucidation of the corresponding classical and quantum dynamics is very involved. It is also true that this type of systems does not have a Lagrangian formulation \cite{Novikov,Henneaux}\footnote{Of course these systems have a first order Lagrangian formulation \cite{HU} that is equivalent to the generalized Hamiltonian dynamics alluded above. Noncommutative dynamical systems does not have a variational formulation in configuration space for Lagrangians of the form $L(\dot q,q)$.}, and basic questions such as what {\em is} the configuration space, what are the symmetries and the variational principle, how the symplectic  noncommutativity manifest itself in configuration space are very difficult to answer. Indeed, we can deduce from the recent analysis of diffeomorphism invariance in noncommutative theories \cite{TS}
that the implementation of symmetries in these theories are difficult to understand at a basic level.

For this class of systems the formulation in phase space is not equivalent to its formulation in configuration space, essentially because the momenta are not auxiliary variables in the first order formulation of the dynamics.

The aim of the present note is to extend our previous investigation \cite{CG1} where we have studied the dynamical compatibility between a Poisson bracket and a given set of second order equations of motion. We have found that if the Poisson bracket is noncommutative 
\begin{equation}\label{npb}
\{q^i,q^j\}=\theta^{ij},
\end{equation}
where $\theta^{ij}$ is a constant antisymmetric matrix,
the Noncommutative Consistency Conditions (NCC), that come from the analysis of the dynamical compatibility, are not the Helmholtz conditions of the generalized inverse problem of the calculus of variations. When $\theta^{ij}=0$ we recover the result that the consistency conditions are the Helmholtz conditions for the given system of second order differential equations \cite{Hojman-Shepley}.

A fundamental question that we have left open in our previous work was if   the NCC can be seen as a $\theta$--deformed Helmholtz conditions associated to some system of differential equations. But if that were the case, the question is which are the differential equations underlying this $\theta$--deformed Helmholtz conditions. Here we will present an answer to this question. Surprisingly enough we found that the differential equations are of {\em third order in time derivatives}.
This could become in a very bad news in the sense that our system --the second order differential equations from we started the analysis of the dynamical compatibility-- is not of this form. Third order dynamical systems lay outside the standard systems studied in Newtonian mechanics. But the third order system that we have found has a very peculiar functional form. It is a functional combination of the original Newtonian equations of motion and its derivatives with respect to time. The general form of a third order system of differential equations that admits in their solution space a subspace that consist of all the paths that are solutions to the original Newtonian problem must be of this general form 
\begin{equation}\label{tode}
N_{ij}(\ddot q^j-F^j(q,\dot q))^{\cdot}+M_{ij}(\ddot q^j-F^j(q,\dot q))=0,
\end{equation}
where $N$ and $M$ are some matrices related to the presence of the noncommutative  parameter $\theta$. In particular, we will require that when $\theta$ goes to zero $N\to 0$, recovering a set of differential equations equivalent to the Newtonian equations of motion. $F^j$ are the forces that define the original Newtonian dynamical system. In this letter we will not be interested in the complete space of solutions to this third order system and we will deal only with the subspace of solutions compatible with the extended Newtonian vector flow given by
\begin{equation}\label{nvf}
\frac{D}{Dt}={\dot F}^i\frac{\partial}{\partial{\ddot q}^i} +
{F}^i\frac{\partial}{\partial{\dot q}^i}+
{\dot q}^i\frac{\partial}{\partial q^i}+
\frac{\partial}{\partial t},
\end{equation}
{\em i.e.,} the vector flow along the solutions of the original system of second order equations of motion and its derivatives with respect to time \footnote{Notice that we are replacing $\ddot q^i=F^i$ and $\dddot q^i=\dot F^i$. It is understood that when we apply this operator to a function of $\ddot q, \dot q$ and $q$ we need to project the final result on $\ddot q^i=F^i$.}. 

Under these assumptions we will present in section 3 our central result that the NCC of our previous work \cite{CG1} are indeed the Helmholtz conditions associated with a system of the form (\ref{tode}). We will find also the relation between the matrices $N$ and $M$ and the symplectic structure
\begin{equation}\label{nss}
\sigma=\left(\begin{array}{cc}B_{ij} & A_{ij} \\-A_{ji} & C_{ij}\end{array}\right),
\end{equation}
and in section 4 we will solve the Helmholtz condition for a generic class of noncommutative systems that can be constructed from a potential function that depends only on the coordinates of the configuration space. Then we will proceed  to construct a Lagrangian in configuration space and the corresponding variational principle for this generic solution. It turns out that the Lagrangian is a linear function in the accelerations --because the equations of motion are of third order--\footnote{Accelerations dependent Lagrangians also appear in the context of the so called Exotic Galilean Symmetries. See for example \cite{Stichel}}.   We will also comment about the relevance of the variational formulation and its possible physical content and quantization. 

Section 2 is devoted to a review of the construction of the NCC between a given set of equations of motion and a Noncommutative Poisson Bracket (NPB). In section 5 we present the Noether theorem and the explicit form of the Lagrangian for the central potential. Finally we present our conclusions and some perspectives for future work.

The third order dynamical system of the form (\ref{tode}) is not new and  was reported in \cite{Hojman} in the context of the study of non-Noetherian symmetries and their relation with the symmetries of the equations of motion. We hope that the analysis presented here can be useful to elucidate many points that were left open in \cite{Hojman} and in particular the relation between noncommutative Poisson structures and variational principles from one hand and from the other the relation between these noncommutative structures in phase space and their consequences for a basic definition of the corresponding noncommutative configuration space.

\section{Consistency Conditions for Noncommutative Poisson Brackets}
In this section we will collect some of the basic results from our previous work that will be useful in what follows.
Given a set of equations of motion in configuration space 
\begin{equation}
\label{eq-m}
\ddot q^i - F^i(q^j,\dot q^j,t)=0,
\end{equation}
and the Poisson bracket defined by
\begin{equation}
\label{NCR}
\{q^i,\dot q^j\}=g^{ij}(q,\dot q),\qquad \{q^i,q^j\}=\theta^{ij},
\end{equation}
where $g^{ij}$ is a symmetric matrix and $\theta^{ij}$ an antisymmetric constant matrix, we say that the NPB (\ref{NCR}) is compatible with the given equations of motion (\ref{eq-m}) if using
\begin{equation}
\label{leibniz}
{\frac{D}{Dt}}\{R,S\}=\{{\frac{D}{Dt}}R,S\} + \{R,\frac{D}{Dt}S\},
\end{equation}
where
$${\frac{D}{Dt}}=F^i\frac{\partial}{\partial\dot q ^i}+\dot q^i\frac{\partial}{\partial q^i}+\frac{\partial}{\partial t},$$
we can construct a matrix $g^{ij}$ and a matrix $b^{ij}$ defined by 
\begin{equation}\label{b-def}
b^{ij}=\{\dot q^i,\dot q^j\},
\end{equation} 
that depends on $g^{ij},\theta^{ij}$ and $F^i$ in such a way
that the corresponding Poisson matrix $\sigma^{ab}$ is consistent with the Leibnitz rule (\ref{leibniz}) and the Jacobi identity \cite{Feynman}. 

Our task is now to find these consistency conditions when $\theta^{ij}$ is different from zero. A straightforward calculation gives
\begin{subequations}
\label{helm}
\begin{align}
\{q^i,g^{jk}\}=\{q^j,g^{ik}\},\label{h-a}\\
\frac{D}{Dt}g^{ij}=\frac12 \{q^i,F^j\}+\frac12 \{q^j,F^i\},\label{h-b}\\
\frac{D}{Dt}b^{ij}= \{\dot q^i,F^j\}- \{\dot q^j,F^i\},\label{h-c}
\end{align}
\end{subequations}
where 
\begin{multline}
\label{B}
b^{ij}=\{\dot q^i,\dot q^j\}= -\frac12 \{q^i,F^j\}+\frac12 \{q^j,F^i\}\\
=\frac12
(\theta^{jk}\frac{\partial F^i}{\partial q^k} - \theta^{ik}\frac{\partial F^j}{\partial q^k}) +\frac12( g^{jk}\frac{\partial F^i}{\partial\dot q^k}- g^{ik}\frac{\partial F^j}{\partial\dot q^k}),
\end{multline}
and $g^{ij}$ a symmetric matrix (this is a simple consequence of $\frac{D}{Dt}
\{q^i,q^j\}=0)$. The first set of equations (\ref{h-a}) comes from the Jacobi identity $\{q^i,\{q^j,\dot q^k\}\} +(ijk)=0$. The equations (\ref{h-b}) come from the derivative of the first bracket in (\ref{NCR}) symmetrizing and adding the two equations.  The equations (\ref{h-c}) come from the derivative of the definition of the matrix $b$. The matrix $b$ in (\ref{B}) can be constructed from the equations (\ref{h-b}) and using the definition of the matrix $g^{ij}$.

We can write these conditions as the Helmholtz conditions plus terms that depend on the noncommutative parameter $\theta$
\begin{subequations}
\label{helmcond}
\begin{align}
&g_{ij}=g_{ji}, \label{hc-a}\\
&\frac{\partial g_{ij}}{\partial\dot{q}^k}-\frac{\partial g_{ik}}{\partial\dot{q}^j}+L^\theta_{ijk}=0, \label{hc-b}\\
&\frac{D}{Dt}g_{ij}=-\frac{1}{2}\left(g_{ik}\frac{\partial F^k}{\partial\dot{q}^j}+g_{jk}\frac{\partial F^k}{\partial\dot{q}^i}\right)+M^\theta_{ij}, \label{hc-c}\\
&\frac{D}{Dt}t_{ij}= g_{ik}\frac{\partial F^k}{\partial q^j}-g_{jk}\frac{\partial F^k}{\partial q^i}+N^\theta_{ij},\label{hc-d}
\end{align}
\end{subequations} 
where
$$t_{ij}=\frac12(g_{ik}\frac{\partial F^k}{\partial\dot{q}^j}-g_{jk}\frac{\partial F^k}{\partial\dot{q}^i}),$$
and $L^\theta_{ijk}, M^\theta_{ij}, N^\theta_{ij}$ are terms that depend on $\theta$. Explicitly  they  are given by
$$L^\theta_{ijk}=\theta^{sl}g_{sk}\frac{\partial g_{ij}}{\partial q^l}- \theta^{rl}g_{rj}\frac{\partial g_{ki}}{\partial q^l},$$
$$M^\theta_{ij}=-\frac12 g_{ir}(\theta^{rn}\frac{\partial F^{s}}{\partial q^n}+\theta^{sn}\frac{\partial F^{r}}{\partial q^n})g_{js},$$
$$N^\theta_{ij}=g_{li}g_{mj}(-\frac{D}{Dt}s^{lm}+s^{lk}\frac{\partial F^m}{\partial\dot{q}^k}-s^{mk}\frac{\partial F^l}{\partial\dot{q}^k})+ g^{lk}(t_{lj}M^\theta_{ik}-t_{li}M^\theta_{jk}),$$
where
$$s^{ij}=\frac12
(\theta^{jk}\frac{\partial F^i}{\partial q^k} - \theta^{ik}\frac{\partial F^j}{\partial q^k}).$$
Here $g_{ij}$ denotes the inverse matrix of our previous $g^{ij}$.

In this letter we will refer to the relations (\ref{helmcond}) as the NCC. They are a set of partial differential equations for $g^{ij}$ given the Newtonian force vector $F^j$ and the noncommutative parameter $\theta$ defined in (\ref{npb}).

These relations are not the Helmholtz conditions associated to the given system of differential equations (\ref{eq-m}) or an s-equivalent related system. 
Even if they are satisfied for some NPB we can not infer the existence of a Lagrangian function of the form $L(q,\dot q)$ associated to the original Newtonian system (\ref{eq-m}). It is known that when the Poisson bracket is noncommutative in the sense of (\ref{NCR}) then a Lagrangian for the system (\ref{eq-m}) can not be constructed \cite{Novikov, Henneaux}. 

In the next section we will use the standard form of the Helmholtz conditions for a system of differential equations of fourth order and apply them to the system (\ref{tode})  and we will compare our result with the symplectic version of the NCC (\ref{helmcond}).

\section{Symplectic formulation of the NCC and Helmholtz Conditions}

Notice that the condition (\ref{leibniz}) is not the standard Leibnitz rule. In fact it implies a dynamical compatibility between the bracket and the dynamical vector field along the solution curves of the equations of motion. If we define the bracket 
by
\begin{equation}\label{PB}
\{R,S\}=\frac{\partial R}{\partial z^a}\sigma^{ab}\frac{\partial S}{\partial z^b},
\end{equation}
where $z^a=(q^i,\dot q^j)$ then the condition (\ref{leibniz}) imply
\begin{equation}\label{helm-cond}
{\cal L}_F(\sigma^{ab})=0,
\end{equation}
where ${\cal L}_F$ is the Lie derivative along the solution vector field associated with the system (\ref{eq-m}). The content of this condition are the Helmholtz conditions. In particular when $\sigma^{ab}$ is given by (\ref{NCR},\ref{b-def}) these relations gives the NCC (\ref{helmcond}). For the purpose of this section it is much more convenient to work with the equivalent form of the NCC associated to the inverse two form $\sigma_{ab}$. The formulation of the conditions (\ref{helm-cond}) in terms of this inverse two-form (the symplectic form $\sigma$) is 
\begin{equation}\label{symp-cond}
{\cal L}_F(\sigma_{ab})=0.
\end{equation}
If we denote the components of the symplectic form by
\begin{equation}\label{symp-form}
\sigma=\left(\begin{array}{cc}B_{ij} & A_{ij} \\-A_{ji} & C_{ij}\end{array}\right),
\end{equation}
the symplectic--NCC are
\begin{subequations}
\label{shc}
\begin{align}\label{shc-a}
&\frac{D}{Dt}A^S_{ij}+A_{ik}\frac{\partial F^k}{\partial {\dot q}^j}
+A_{jk}\frac{\partial F^k}{\partial {\dot q}^i}
-C_{jk}\frac{\partial F^k}{\partial {q}^i}
-C_{ik}\frac{\partial F^k}{\partial {q}^j}=0,\\ \label{shc-b}
&\frac{D}{Dt}A^A_{ij}+A_{ik}\frac{\partial F^k}{\partial {\dot q}^j}-
A_{jk}\frac{\partial F^k}{\partial {\dot q}^i}
+C_{ik}\frac{\partial F^k}{\partial {q}^j}
-C_{jk}\frac{\partial F^k}{\partial {q}^i}+2 B_{ij}=0,\\ \label{shc-c}
&\frac{D}{Dt}B_{ij}+A_{ik}\frac{\partial F^k}{\partial { q}^j}-A_{jk}\frac{\partial F^k}{\partial {q}^i}=0,\\ \label{shc-d}
&\frac{D}{Dt}C_{ij}+C_{ik}\frac{\partial F^k}{\partial {\dot q}^j}-C_{jk}\frac{\partial F^k}{\partial {\dot q}^i}+A^A_{ij}=0,
\end{align}
\end{subequations}
where $B$ and $C$ are antisymmetric matrices and $A^{(A)S}$ is the (anti)symmetric part of $A$ respectively which are defined by $A^A=A-A^T$ and $A^S=A+A^T$.
 
 For our purpose it is convenient to eliminate $B$ using (\ref{shc-b}) 
 and (\ref{shc-d}). The symplectic--NCC are then
  \begin{subequations}
\label{shc1}
\begin{align}\label{shc-a1}
&\frac{D}{Dt}A^S_{ij}+A_{ik}\frac{\partial F^k}{\partial {\dot q}^j}
+A_{jk}\frac{\partial F^k}{\partial {\dot q}^i}
-C_{jk}\frac{\partial F^k}{\partial {q}^i}
-C_{ik}\frac{\partial F^k}{\partial {q}^j}=0,\\ \label{shc-b1}
&\frac{D}{Dt}B_{ij}(A,C,F)+A_{ik}\frac{\partial F^k}{\partial { q}^j}-A_{jk}\frac{\partial F^k}{\partial {q}^i}=0,\\ \label{shc-c1}
&\frac{D}{Dt}C_{ij}+C_{ik}\frac{\partial F^k}{\partial {\dot q}^j}-C_{jk}\frac{\partial F^k}{\partial {\dot q}^i}+A^A_{ij}=0,
\end{align}
\end{subequations}
where $B(A,C,F)$ is given by
 \begin{multline}\label{B-def}
 B_{ij}(A,C,F)=-\frac12\left(-\frac{D^2}{Dt^2}C_{ij}+\frac{D}{Dt}\left(C_{jk}\frac{\partial F^k}{\partial\dot q^i}\right)-\frac{D}{Dt}\left(C_{ik}\frac{\partial F^k}{\partial\dot q^j}\right)\right.\\ 
 +A_{ik}\frac{\partial F^k}{\partial\dot q^j}-
 \left . A_{jk}\frac{\partial F^k}{\partial\dot q^i}- C_{jk}\frac{\partial F^k}{\partial q^i}+
 C_{ik}\frac{\partial F^k}{\partial q^j}\right),
 \end{multline}

Notice that we need to solve the symplectic--NCC only for $A$ and $C$ for a given $F$. The matrix $B$ can then be calculated from (\ref{B-def}).
 
 In the following we will not need the explicit form of the matrices $A,B,C$ in terms of the NPB matrices $g,b$ and $\theta$. It is sufficient to keep in mind that they are functions of $q,\dot q$ and $\theta$ and can be obtained from the NPB through $\sigma^{ab}\sigma_{bc}=\delta^a_c$.
 
Now the Helmholtz conditions associated to a system of differential equations of fourth order namely those systems that come from a variational principle whose Lagrangian is a function of positions, velocities and accelerations are \cite{Anderson}
\begin{subequations}
\label{HC-4}
\begin{align}\label{HC-a}
&\frac{\partial {\cal M}_i}{\partial\ddddot q^k}=\frac{\partial {\cal M}_k}{\partial\ddddot q^i},\\ \label{HC-b}
&\frac{\partial {\cal M}_i}{\partial\dddot q^k}+\frac{\partial {\cal M}_k}{\partial\dddot q^i}=2\frac{{\cal D}}{{\cal D}t}\left(\frac{\partial {\cal M}_i}{\partial\ddddot q^k}+\frac{\partial {\cal M}_k}{\partial\ddddot q^i}\right),\\ \label{HC-c}
&\frac{\partial {\cal M}_i}{\partial\ddot q^k}-\frac{\partial {\cal M}_k}{\partial\ddot q^i}=\frac{3}{2}\frac{{\cal D}}{{\cal D}t}\left(\frac{\partial {\cal M}_i}{\partial\dddot q^k}-\frac{\partial {\cal M}_k}{\partial\dddot q^i}\right),\\ \label{HC-d}
&\frac{\partial {\cal M}_i}{\partial\dot q^k}+\frac{\partial {\cal M}_k}{\partial\dot q^i}=\frac{{\cal D}}{{\cal D}t}\left(\frac{\partial {\cal M}_i}{\partial\ddot q^k}+\frac{\partial {\cal M}_k}{\partial\ddot q^i}\right)- \frac{{\cal D}^3}{{\cal D}t^3}\left(\frac{\partial {\cal M}_i}{\partial\ddddot q^k}+\frac{\partial {\cal M}_k}{\partial\ddddot q^i}\right),\\ \label{HC-e}
&\frac{\partial {\cal M}_i}{\partial q^k}-\frac{\partial {\cal M}_k}{\partial q^i}=\frac12\frac{{\cal D}}{{\cal D}t}\left(\frac{\partial {\cal M}_i}{\partial\dot q^k}-\frac{\partial {\cal M}_k}{\partial\dot q^i}\right)-\frac14 \frac{{\cal D}^3}{{\cal D}t^3}\left(\frac{\partial {\cal M}_i}{\partial\dddot q^k}+\frac{\partial {\cal M}_k}{\partial\dddot q^i}\right),
\end{align}
\end{subequations}
where
\begin{equation}\label{diff-eq}
{\cal M}_i(\ddddot q^j,\dddot q^j,\ddot q^j,\dot q^j,q^j)=0,
\end{equation}
is the the set of differential equations under consideration and
$\frac{{\cal D}}{{\cal D}t}$
is the time derivative along the solutions of the system ${\cal M}_i=0$. Implementing these Helmholtz conditions for the class of systems that are  under our consideration (\ref{tode}) along the Newtonian vector flux given by (\ref{nvf}) and identifying 
$$\frac{{\cal D}}{{\cal D}t}\to \frac{{ D}}{{ D}t},$$
and the matrices $N$ and $M$ with
$$N_{ij}=C_{ij},\qquad M_{ij}=-A_{ij}+\frac{d}{dt}C_{ij},$$
where $A$ and $C$ are defined through the symplectic form (\ref{symp-form}), we will prove that these Helmholtz conditions (\ref{HC-4}) are the same as the symplectic-NCC (\ref{shc1}). 

The first condition (\ref{HC-a}) is in our case trivial.

From the second condition (\ref{HC-b}) we have
$$C_{ik}+C_{ki}=0,$$
that just confirm that $C$ is antisymmetric.

From the third condition (\ref{HC-c}) we have
\begin{equation}\label{shc-d-bis}
\frac{D}{ D t} C_{ik}=C_{kj}\frac{\partial F^j}{\partial\dot q^i}-C_{ij}\frac{\partial F^j}{\partial\dot q^k}-A^A_{ik},
\end{equation}
which coincide exactly with the symplectic--NCC (\ref{shc-c1}).

From the fourth condition (\ref{HC-d}) we have
\begin{eqnarray}\nonumber
&&\frac{D}{ D t}\left( A^S_{ik} + C_{ij}\frac{\partial F^j}{\partial\dot q^k}+C_{kj}\frac{\partial F^j}{\partial\dot q^i}\right)\\ \nonumber
&=&C_{ij}\frac{\partial\dot F^j}{\partial\dot q^k}+C_{kj}\frac{\partial\dot F^j}{\partial\dot q^i}+(-A_{ij}+\frac{D}{Dt}C_{ij})\frac{\partial F^j}{\partial\dot q^k}+(-A_{kj}+\frac{D}{Dt}C_{kj})\frac{\partial F^j}{\partial\dot q^i}.
\end{eqnarray}
Now, using\footnote{Notice that the dots over the vector $F$ denote standard derivatives with respect to time. We are allowed to evaluate these relations along the Newtonian vector flux (\ref{nvf}) at the end of our calculation.}
$$\frac{\partial}{\partial\dot q^k}\frac{d}{dt}-\frac{d}{dt}\frac{\partial}{\partial\dot q^k}=\frac{\partial}{\partial q^k},$$
and evaluating  the expression over the vector field (\ref{nvf}) we obtain the symplectic--NCC (\ref{shc-a1}),
\begin{equation}\label{shc-b-bis}
\frac{D}{ D t} A^S_{ik}=C_{ij}\frac{\partial F^j}{\partial\dot q^k}+C_{kj}\frac{\partial F^j}{\partial\dot q^i}-A_{ij}\frac{\partial F^j}{\partial\dot q^k}-A_{kj}\frac{\partial F^j}{\partial\dot q^i}.
\end{equation}
From the last Helmholtz condition (\ref{HC-e}) we find
\begin{eqnarray}\nonumber
&\frac12\frac{D}{Dt}\left(-C_{ij}\frac{\partial\dot F^j}{\partial\dot q^k}+C_{kj}\frac{\partial\dot F^j}{\partial\dot q^i}-(-A_{ij}+\frac{D}{Dt}C_{ij})\frac{\partial F^j}{\partial\dot q^k}+(-A_{kj}+\frac{D}{Dt}C_{kj})\frac{\partial F^j}{\partial\dot q^i}  -\frac{D^2}{Dt^2}C_{ik}\right)\\ \nonumber &
=-C_{ij}\frac{\partial\dot F^j}{\partial q^k}+C_{kj}\frac{\partial\dot F^j}{\partial q^i}-(-A_{ij}+\frac{D}{Dt}C_{ij})\frac{\partial F^j}{\partial q^k}+(-A_{kj}+\frac{D}{Dt}C_{kj})\frac{\partial F^j}{\partial q^i}.
\end{eqnarray}
This relation can be reduced to the corresponding 
symplectic--NCC (\ref{shc-b1}) 
\begin{equation}\label{shc-c-bis}
\frac{D}{Dt}B_{ij}(A,C,F)+A_{ik}\frac{\partial F^k}{\partial { q}^j}-A_{jk}\frac{\partial F^k}{\partial {q}^i}=0,
\end{equation}
where $B(A,C,F)$ is given by (\ref{B-def}).

The conditions (\ref{shc-d-bis},\ref{shc-b-bis},\ref{shc-c-bis}) are the same as the symplectic NCC (\ref{shc1}). Thus we have proved that the NCC can be seen as  $\theta$--deformed Helmholtz conditions under the basic assumptions

1) The equation of motion have the general form
\begin{equation}\label{tode1}
N_{ij}(\ddot q^j-F^j(q,\dot q))^{\cdot}+M_{ij}(\ddot q^j-F^j(q,\dot q))=0,
\end{equation}

2) The vector field along the solution curves of this system is projected to 
$$\frac{{\cal D}}{{\cal D}t}\to \frac{{ D}}{{ D}t},$$

where
\begin{equation}
\frac{D}{Dt}={\dot F}^i\frac{\partial}{\partial{\ddot q}^i} +
{F}^i\frac{\partial}{\partial{\dot q}^i}+
{\dot q}^i\frac{\partial}{\partial q^i}+
\frac{\partial}{\partial t}.
\end{equation}

3) The identification
\begin{equation}\label{iden}
N_{ij}=C_{ij},\qquad M_{ij}=-A_{ij}+\frac{d}{dt}C_{ij},
\end{equation}

where $A$ and $C$ are defined through the symplectic form (\ref{symp-form}).

In summary we conclude that the $\theta$-deformed Helmholtz conditions (\ref{shc1}) are in fact the Helmholtz conditions associated to a $\theta$-deformed Newtonian system (\ref{tode1}) where $N$ and $M$ are related with the symplectic form $\sigma$ by the identification (\ref{iden}).  

As this is the central result of this letter let us comment about some physical consequences of it.

$\bullet$ Starting from the dynamical system defined by
\begin{equation}\label{tode2}
C_{ij}(\ddot q^j-F^j(q,\dot q))^{\cdot}+(-A_{ij}+\frac{d}{dt}C_{ij})(\ddot q^j-F^j(q,\dot q))=0,
\end{equation}
and considering the sector of the solution space associated to this system that is compatible with the Newtonian vector flux
$$\frac{D}{Dt}={\dot F}^i\frac{\partial}{\partial{\ddot q}^i} +
{F}^i\frac{\partial}{\partial{\dot q}^i}+
{\dot q}^i\frac{\partial}{\partial q^i}+
\frac{\partial}{\partial t},$$
we have found that the Helmholtz conditions (\ref{HC-4}) associated  to it are the compatibility consistency conditions between the Newtonian  system (\ref{eq-m}) and the noncommutative Poisson structure (\ref{NCR}). This imply that, in noncommutative space,  the only way in which a Newtonian system can be brought to a variational formulation\footnote{Up to some simple cases that was analyzed in \cite{CG1}.} is by the use of an auxiliary third order system of equations of motion (\ref{tode2}) which have the peculiar characteristic of being a functional combination of the original Newtonian equations of motion and its derivatives with respect to time. The coefficients of this functional combinations are completely fixed and given by the noncommutative symplectic structure (\ref{symp-form}).

$\bullet$ The relevant paths for the variational formulation are solutions to the original Newtonian system. The other solutions of the third order system (\ref{tode2}) are not considered in the formulation of our compatibility problem. We are just asking about the fate of a Newtonian system formulated in a noncommutative space. We can make a wide question asking for the class of systems with higher derivatives that allow a noncommutative formulation. In that case the complete third order problem may have some relevance. This type of systems lay outside of the Newtonian paradigm.

$\bullet$ As we said before it is known that, in general, the Newton equations does not admit a variational formulation in a noncommutative space.

$\bullet$ Perhaps is not too surprising that if we insist in a variational formulation of a system that does not admit a Lagrangian of the form $L(\dot q,q)$ then  a new ($\theta$-deformed) auxiliary system of differential equations (\ref{tode2}) emerge allowing a variational formulation. 

\section{Generic solution to the Helmholtz Conditions}

In our previous work we have presented a generic solution
of the NCC (\ref{helmcond}) for forces of the form
\begin{equation}
\label{nc-f}
F^i=-\frac{\partial V}{\partial q^i}+\theta^{ij}\frac{d}{dt}\frac{\partial V}{\partial q^j},
\end{equation}
for a ``potential function'' $V(q)$. This type of forces arise in the first order formulation for a Hamiltonian function of the form $H=\frac12 p^2+V$ with the symplectic structure
\begin{equation}
\label{sym-xp}
\{q^i,q^j\}=\theta^{ij},\quad \{p_i,p_j\}=0,\quad \{q^i,p_j\}=\delta^i_j,
\end{equation}
which is called in current literature of noncommutative physics as a noncommutative dynamical system.
The Hamiltonian equations of motion associated with  this system are
\begin{subequations}
\label{h-eq}
\begin{align}
\dot q^i + \theta^{ij}\dot{p_j}- p_i=0,\label{he-a}\\
-\dot{p_i} -\frac{\partial V}{\partial q^i}=0.\label{he-b}
\end{align}
\end{subequations}
Forces in configuration space of the form (\ref{nc-f}) can be deduced from these first order equations by using the set (\ref{he-b}) and taking the derivative with respect to time of the first set of equations (\ref{he-a}). 

The NPB associated to the forces (\ref{nc-f}) that solve the $\theta$-deformed Helmholtz conditions (\ref{helmcond}) are
\begin{subequations}
\label{nc-NCR}
\begin{align}
\{q^i,\dot q^j\}=\delta^i_j+V^{ij},\quad \{q^i,q^j\}=\theta^{ij},\label{nc-NCR-a}\\
\{\dot q^i,\dot q^j\}=V^i_{j}+\frac12 V^i_{n} V^{jn}-(i\leftrightarrow j),\label{nc-NCR-b}
\end{align}
\end{subequations}
where we are using the notation
$$V^i\equiv\theta^{ij}V_j=\theta^{ij}\frac{\partial V}{\partial q^j}$$
and the associated equations of motion that are compatible with this NPB are
\begin{equation}
\label{nc-eq-m}
\ddot q^i+ V_i-\dot V^i=0. 
\end{equation}
This imply that the classical system defined by the equations of motion (\ref{nc-eq-m}) and the NPB (\ref{nc-NCR}) are dynamically compatible.

Inspired from this solution to the $\theta$-deformed Helmholtz conditions (\ref{helmcond}) we can guess the corresponding solution of the symplectic-NCC (\ref{shc1}) along the vector field generated by the forces (\ref{nc-f}). The solution is
\begin{subequations}
\label{nc-symp}
\begin{align}\label{nc-A}
&A_{ij}=-\delta_{ij}+\theta^{jk}V^k_i,\\ \label{nc-B}
&B_{ij}=V^i_j-V^j_i-\theta^{kl}V^l_{i}V^k_{j},\\ \label{nc-C}
&C_{ij}=\theta^{ij}.
\end{align}
\end{subequations}

It is easy to check that the corresponding symplectic form is the inverse of the Poisson bracket given by (\ref{nc-NCR}) as it should be. The next step is to construct a Lagrangian associated to this symplectic form (\ref{nc-symp}). Starting from the most general Lagrangian function linear in $\ddot q^i$ we can construct the equations of motion and compare the result with the system of third order differential equations (\ref{tode2}) associated with the symplectic form given above (\ref{nc-symp}). As a result we can read off a Lagrangian whose equations of motion are of the form (\ref{tode2}) for the specific functional form of the forces (\ref{nc-f}) associated with the noncommutative dynamical system whose potential function is $V(q)$. The resulting Lagrangian linear in $\ddot q$ is
\begin{equation}\label{nc-L}
L(\ddot q,\dot q,q)=L_0-\frac12\theta^{ij}\dot q^i \dot V^j+\frac12\theta^{ij}    
\ddot q^j(\dot q^i-V^i)+\frac12\theta^{ij}V^i\dot V^j-\frac12 V^kV^k,
\end{equation}
where $L_0$ is the standard commutative Lagrangian $L_0=T-V$.
By construction the Lagrangian (\ref{nc-L}) still admits a symplectic formulation. Regarding $q,\dot q$ as independent variables, as is usual in the standard commutative case and $\ddot q$ as a derived quantity that is just the derivative with respect to time of $\dot q$, the symplectic formulation of the Lagrangian (\ref{nc-L}) is
$$L=\ell_i(q,\dot q)\dot q^i + {\cal L}_i(q,\dot q)\ddot q^i +{\cal V}(q),$$ 
where
$$\ell_i=\frac12\dot q^i-\frac12\theta^{ij}\dot V^j+\frac12\theta^{kj}V^k V_i^j,$$
$${\cal L}_i=\frac12\theta^{ki} (\dot q^k-V^k),$$
$${\cal V}=-V-\frac12 V^kV^k.$$
From here we can read the symplectic form $\sigma$
$$A_{ij}=\frac{\partial\ell_i}{\partial \dot q^j}-\frac{\partial{\cal L}_j}{\partial  q^i}=-\delta_{ij}+\theta^{jk}V^k_i,$$
$$B_{ij}=\frac{\partial\ell_i}{\partial q^j}-\frac{\partial\ell_j}{\partial  q^i}=V^i_j-V^j_i-\theta^{kl}V^l_{i}V^k_{j},$$
$$C_{ij}=\frac{\partial{\cal L}_i}{\partial \dot q^j}-\frac{\partial{\cal L}_j}{\partial \dot q^i}=\theta^{ij},$$
that coincide with (\ref{nc-symp})  as it should be. Notice that we have a lot of freedom in the definition of $\ell_i,{\cal L}_i$. This freedom comes from the fact that $\sigma$ is a two-form that comes from the one-form defined by $\ell_i,{\cal L}_i$.

The corresponding variational principle is
\begin{equation}\label{nc-S}
S=\int_{t_1}^{t_2}L(\ddot q,\dot q,q) dt,
\end{equation}
and its associated equations of motion are the auxiliary system of third order differential equations (\ref{tode2}). This action can be used to construct integrals of motion for each Noetherian symmetry. Formally we can use the Weiss action principle \cite{Sudarshan} to obtain from (\ref{nc-S}) the equations of motion (\ref{tode2})
$$\delta S[q(t)]=G(t_2)-G(t_1)\to C_{ij}(\ddot q^j-F^j)^{\cdot}-A_{ij}(\ddot q^j-F^j)=0,$$
where $F^j$, $A_{ij}$ and $C_{ij}$ are defined in (\ref{nc-f}),(\ref{nc-A}) and (\ref{nc-C}) respectively.
We need to keep in mind that we are interested in the dynamics of the Newtonian system
$$(\ddot q^j-F^j)=0,$$
and not on the dynamics of the auxiliary system (\ref{tode2}). The solution space of the Newtonian system is a subsector of all the solutions of the system (\ref{tode2}).

We will end this section with a comment on the question if the action (\ref{nc-S}) 
can be used as a starting point to quantize the noncommutative dynamical system via the Feynman Path Integral approach. As is well known, this quantization method is consistent provided 
\begin{equation}\label{var-pri}
\delta S[q(t)]=0,
\end{equation}
for the class of paths under consideration. 
To obtain (\ref{tode2}) upon variation of (\ref{nc-S}) we need to supplement the action with some boundary terms.
Notice that this point is crucial because our variational principle have at least two extremals. One corresponding to the solutions of the third order equations of motion and the other corresponding to the solutions of the second order Newtonian equations of motion (\ref{eq-m}).
As in the commutative case any set of variables that Poisson commute at the time $t_1$ and $t_2$ are good variables to fix at boundaries. It is usual to pick the positions at the time 1 and 2 to obtain $\langle1|2\rangle$ in the Heisenberg picture. In the noncommutative case we have less freedom to choose the boundary. In our case a natural choice is\footnote{These variables correspond to the momenta in associated first order formulation.}
\begin{equation}\label{b-data}
(\dot q^i-V^i)(t_1)=Q^i(t_1),\quad (\dot q^i-V^i)(t_2)=Q^i(t_2),\quad \delta Q^i(t_1)=\delta Q^i(t_2)=0.
\end{equation}
With this choice we have now control over the variations of the action (\ref{nc-S}) by specifying the correct dynamical information to recover the Newtonian dynamics from it.

Indeed, a general variation of the action (\ref{nc-S}) produces the boundary term
$$\delta S=\int^{t_2}_{t_1}EL_i(L)\delta q^i+\left(\frac{\partial L}{\partial\ddot q^i}\delta \dot q^i+\frac{\partial L}{\partial \dot q^i}\delta q^i-\frac{d}{dt}\left(\frac{\partial L}{\partial\ddot q^i}\right)\delta q^i+\delta B\right)\Bigg|_{t_1}^{t_2},$$
that in our specific case gives $(B=-\delta(q^j Q^j))$ 
$$\left(\frac12 \theta^{kj}(\dot q^k-V^k)-q^j\right)\delta Q^j-\left(\theta^{kj}(\ddot q^k-\dot V^k+V_k)\right)\delta q^j.$$
These boundary terms are  zero if we take into account (\ref{b-data}) {\em and} enforce by hand the second order Newtonian equations of motion to zero at $t_1$ and $t_2$
\footnote{Of course it is also possible to choose the variables $Q^j=\frac12 \theta^{kj}(\dot q^k-V^k)-q^j$ to fix at the boundaries. They corresponds to the Darboux transformation used to get the commutative variables from the noncommutative ones in the first order formulation.}.
As we need to enforce the Newtonian equations of motion at the boundaries $t_1$ and $t_2$ in order that the variational principle meets the criteria (\ref{var-pri})  its application for quantization is not yet clear for us. 

We stress that from our point of view --and the only consistent with the dynamical compatibility between the Newtonian equations of motion and the NPB--  the variational principle (\ref{var-pri}) must be formulated in terms of an auxiliary system of differential equations of third order . Even though the global properties of the variational principle are not well understood it can be used to study the relation between symmetries and conserved quantities using an extended version of the Noether theorem. We will see how it works in some simple examples in the next section.

\section{Examples}

In this section we will present some examples of the application of the variational principle that we have constructed on our previous section.

\subsection{Noether theorem}

In general, if the Lagrangian (\ref{nc-L}) is invariant under a Noether symmetry transformation given by $\delta q^i$, then
$$\delta L=\frac{d G}{dt}.$$
From here we can obtain
$$EL_i(L)\delta q^i +\frac{d}{dt}\left(\frac{\partial L}{\partial\ddot q^i}\delta \dot q^i+\frac{\partial L}{\partial \dot q^i}\delta q^i-\frac{d}{dt}\left(\frac{\partial L}{\partial\ddot q^i}\right)\delta q^i- G\right)=0.$$
Now, using the fact we we are interested in the subsector of the solution space of the Euler-Lagrange equations of motion defined by the projection $\ddot q^i=F^i$ and $\dddot q^i=\dot F^i,$
the associated conserved charge is
$$K(q,\dot q,t)=\frac{\partial L}{\partial\ddot q^i} \frac{D}{Dt}\delta q^i+\frac{\partial L}{\partial \dot q^i}\delta q^i-\frac{D}{Dt}\left(\frac{\partial L}{\partial\ddot q^i}\right)\delta q^i-G,$$
where we must keep in mind that any dependence in $K$ on $\ddot q^i$ must be replaced by $F^i$. This procedure is completely consistent and can be very useful to obtain new conservation laws for specific forms of the potential $V$.

Our first example is an application of this extended Noether theorem to obtain the energy conservation from the symmetry associated with time translation when $V$ is independent of $t$. It is well known that this symmetry is $\delta q^i=\dot q^i \delta t$ where $\delta t$ is an infinitesimal parameter. The conserved quantity associated with this symmetry is
$$E=\frac{\partial L}{\partial\ddot q^i}\frac{D}{Dt}\delta q^i+\frac{\partial L}{\partial \dot q^i}\delta q^i-\frac{D}{Dt}\left(\frac{\partial L}{\partial\ddot q^i}\right)\delta q^i-L=\frac12 \dot q^2+V+V^k(\frac12 V^k-\dot q^k).$$
We will call this quantity the ``energy'' for obvious reasons. By the same token we can call ``linear momenta''  the quantity
$$P_i=\frac{\partial L}{\partial \dot q^i}-\frac{D}{Dt}\left(\frac{\partial L}{\partial\ddot q^i}\right).$$
that result from the invariance under translations in the $i$-direction.

\subsection{Central Potential} 

To illustrate a particular form of the Lagrangian (\ref{nc-L}) we will choose a central potential $V(r)$ where $r=(q^i q^i)^{\frac12}$ in $d$ dimensions. The Lagrangian is
$$L=L_0+\frac12 \theta^{ij} \dot q^i \ddot q^j+\theta^{ij}\theta^{jl}\frac{V'}{r}\ddot q^i q^l+\frac12 \theta^{ij}\theta^{jl}\left(\frac{V'}{r}\right)^2\left(q^i q^l-\theta^{ik}\dot q^k q^l\right).$$
where $V'$ is the derivative of $V$ with respect to its argument.
The diverse terms in the Lagrangian are organized in powers of $\theta$. Notice that the free particle still have a dependence linear in $\theta$. This is because this Lagrangian give rise to the allowed third order system of differential equations. An interesting feature of the variational formulation underlaying this Lagrangian prescription is that the corresponding Noether symmetry associated with rotations is now the ``twisted'' symmetry
\begin{equation}\label{symm}
\delta q^i=\dot q^i+\theta_{ij}q^j,
\end{equation}
where $\theta_{ij}$ is the inverse matrix of the matrix $\theta^{ij}$ that we have used along the main text. The associated Noether integral of motion is 
\begin{equation}\label{K}
K=\dot q^i\theta_{ij} q^j-r V'+V.
\end{equation}
This integral of motion is of the form of an ``angular momenta'' of the standard theory of central potential in Classical Mechanics. Notice however, that this conserved quantity is only one and not a vector as in the usual case. 
That (\ref{symm}) is a Noether symmetry is a fact that can only be established with the aid of our Lagrangian formulation.

\section{Conclusions}

In this communication we have presented a solution to the question raised in our previous work \cite{CG1}.  Are the dynamical compatibility conditions between a given Newtonian dynamical system and a NPB the Helmholtz conditions associated to some set of differential equations? The answer is yes but the system of differential equations are of third order in time derivatives. Nevertheless, the particular form of the third order system is such that we can recover from it the dynamical information of the original Newtonian equations of motion. Of course it have more solutions but the form and fate of them are outside the scope of this article.

We can restate our result by saying that the NCC between a NPB and a second order Newtonian dynamical system are the Helmholtz conditions of a $\theta$-deformed system of differential equations (\ref{tode2}).

We have presented a solution to these Helmholtz conditions analogous to the $L=T-V$ prescription of standard Classical Dynamics. The corresponding variational formulation can be used to construct Noether symmetries and conserved quantities that otherwise are much more difficult to unravel.
The relation of the Noether symmetries of this noncommutative Lagrangian and a version of a twisted-Poisson (or twisted-Heisenberg) algebra and the associated implementation of the symmetry concept in this context are actually under investigation.

We have also pointed out some open questions about the usefulness of this variational formulation for the quantization of classical noncommutative systems. We expect to return to the analysis of these important questions in the future. 

Another interesting topic that we are leaving aside in this note is the problem of nonequivalent quantizations of classically equivalent formulations of dynamical systems. For a recent discussion about this topic the reader can consult \cite{Marmo} and in the context of the quantization of String Theory \cite{Thiemann}.

\section*{Acknowledgments} 

This work was supported in part by grants CONACyT 32431-E and DGAPA IN104503. It is a pleasure to thanks P. A. Horvathy for his comments on the relation between first and second order formulations of the dynamics while this work was in progress. We also thanks to Peter Stichel for his comments on the manuscript and to  Simon Lyakhovich and  Alexei Sharapov  for the clarification of the relation of our work with its recent paper \cite{Lyan}.

\end{document}